\documentclass[a4paper,11pt]{article}
\usepackage{amsmath}
\usepackage{amsfonts}
\usepackage{amssymb}
\usepackage[utf8x]{inputenc}
\usepackage{ae}
\usepackage[T1]{fontenc}
\usepackage[english]{babel}
\usepackage{graphicx}
\usepackage{placeins}
\usepackage{tabularx}
\usepackage{tabulary}
\usepackage{setspace}
\usepackage{tablefootnote}
\usepackage[flushmargin]{footmisc}
\usepackage[comma]{natbib}
\usepackage{array}
\usepackage{float}
\usepackage{caption}
\usepackage{tikz}
\usetikzlibrary{shapes.geometric}
\usepackage[a4paper,left=2.5cm,right=2.5cm,top=3cm,bottom=3cm]{geometry}
\usepackage[colorinlistoftodos]{todonotes}
\usepackage{pdfpages}
\usepackage{longtable}
\usepackage{adjustbox}
\usepackage{booktabs}
\usepackage{multirow}
\usepackage{lscape}
\usepackage[colorlinks=true, allcolors=blue]{hyperref}

\usepackage[flushleft]{threeparttable}
\usepackage{inputenc}



\begin{document} \doublespacing \pagestyle{plain}
	
	\def\ci{\perp\!\!\!\perp}
	\begin{center}
		
		{\LARGE Do price reductions attract customers in urban public transport? A synthetic control approach}
		
		{\large \vspace{0.8cm}}
		
		{\large Hannes Wallimann, Kevin Blättler and Widar von Arx }\medskip
		
		{\small {University of Applied Sciences and Arts Lucerne, Institute of Tourism and Mobility} \bigskip }
		
		{\large Version March 2022}
	\end{center}
	
	\smallskip

	\noindent \textbf{Abstract:} {In this paper, we assess the demand effects of lower public transport fares in Geneva, an urban area in Switzerland. Considering a unique sample based on transport companies’ annual reports, we find that, when reducing the costs of annual season tickets, day tickets and hourly tickets (by up to 29\%, 6\% and 20\%, respectively), demand increases by, on average, over five years, about 10.6\%. To the best of our knowledge, we are the first to show how the synthetic control method \citep{abadie2003economic,abadie2010synthetic} can be used to assess such (for policy-makers) important price reduction effects in urban public transport. Furthermore, we propose an aggregate metric that inherits changes in public transport supply (e.g., frequency increases) to assess these demand effects, namely passenger trips per vehicle kilometre. This metric helps us to isolate the impact of price reductions by ensuring that companies' frequency increases do not affect estimators of interest. In addition, we show how to investigate the robustness of results in similar settings. Using a recent statistical method and a different study design, i.e., not blocking off supply changes as an alternate explanation of the effect, leads us to a lower bound of the effect, amounting to an increase of 3.7\%. Finally, as far as we know, it is the first causal estimate of price reduction on urban public transport initiated by direct democracy. 
	}
	
	{\small \smallskip }
	{\small \smallskip }
	{\small \smallskip }
	

	{\small \smallskip }
	{\small \smallskip }
	{\small \smallskip }
	

	{\small \smallskip }
	{\small \smallskip }
	{\small \smallskip }
	
	{\small \noindent \textbf{Acknowledgments:}  The authors would like to thank Helen Conradin, Michael Steinle, Silvio Sticher, Lea Oberholzer, Philipp Wegelin and Martin Huber for helpful comments. Moreover, we are grateful to the SBB Research Fund for financial support.}

	{\small \smallskip }
	{\small \smallskip }
	{\small \smallskip }
	
	{\small \noindent  {\scriptsize 
			\textbf{Addresses for correspondence:} Hannes Wallimann, University of Applied Sciences and Arts Lucerne, Rösslimatte 48, 6002 Lucerne, hannes.wallimann@hslu.ch; Kevin Blättler, kevin.blaettler@hslu.ch; Widar von Arx, widar.vonarx@hslu.ch.
		}\thispagestyle{empty}\pagebreak  }

	{\small \renewcommand{\thefootnote}{\arabic{footnote}} %
		\setcounter{footnote}{0}  \pagebreak \setcounter{footnote}{0} \pagebreak %
		\setcounter{page}{1} }
	
\section{Introduction}\label{introduction}
	
The transport sector is a pivotal contributor to air pollution. Globally, approximately 27\% of CO\textsubscript{2} emissions and energy consumption are caused by the transport sector; in the European Union, the figure amounts to about a third \citep{batty2015challenges}. Therefore, the transport sector causes negative externalities, which means a situation in which the action of a person imposes a cost on another person who is not a party to the transaction. Another important example is noise pollution. Private car use will lead to even greater levels of such negative externalities, which a shift in transport mode towards public transport could help reduce. Lower fares are a frequently discussed tool to motivate individuals to use public transport \citep[see, e.g.,][]{redman2013quality}. 

Policy-makers must know how existing and potential customers respond to such lower fares. However, it is generally challenging to identify the causal effect of lower fares on public transport demand, as transport supply change over time. Therefore, we propose and discuss an aggregate metric that inherits a transport company's supply in public transport demand in this context. The metric is composed of passenger trips per vehicle kilometre. Moreover, considering CO\textsubscript{2} emissions, an increase in the metric points to an average emission decrease of each passenger.

In our comparative case study, we use this metric as the outcome variable to analyze lower fares empirically in the case of Geneva, an urban area in Switzerland. There, the electorate decided to reduce the price of state-owned public transport, which Geneva then introduced in December 2014. The reduction amounted to up to 29\% for annual season tickets, 6\% for day tickets and 20\% for tickets valid for one hour. The case of Geneva is interesting for several reasons. First, Geneva is densely populated. Second, Switzerland has a high per-capita income, as does Geneva. Based on the first and second reasons, we resolve the puzzle of how lower fares cause demand when density and incomes are high, which is the case for many cities worldwide. And third, the public transportation sector in Switzerland is known for its high quality of service. Conclusions can thus also be drawn as to whether price reductions increase the demand for public transport in areas where the quality of the public transport sector is high.

To illustrate the price-reduction effect, we analyze the case of TPG, the main operator in the city of Geneva, and its agglomeration belt. To this end, we apply the synthetic control method \citep{abadie2010synthetic,abadie2003economic} to construct a synthetic TPG, a counterfactual that mimics the demand the company would have experienced in the absence of the price reduction. The methodology uses a data-driven procedure to create the synthetic TPG from comparable Swiss transport operators. Comparing the demand of TPG and its synthetic counterpart, we find that, on average, the price reduction increased the demand for public transport by 10.6\% during the period 2015 to 2019, compared to 2014. 

Furthermore, we set out to block off alternate reasons leading to our estimate through various robustness checks. For example, we find that the results are similar when increasing the length of the pre-treatment period or increasing the number of other operators to construct the synthetic TPG. Moreover, applying the recent difference in differences method of \cite{arkhangelsky2019synthetic} does not question our findings. However, when we set out to assess the effect of our mechanism of interest, the effect of a price reduction on demand, using the total amount of passenger trips instead of the proposed metric, we are not able to construct a suitable synthetic TPG. The thing to notice is that when we re-estimate the effect with the, for this case, more appropriate synthetic difference in the difference method, we only get an effect of 3.7\%. However, a corresponding 95\% bootstrap interval amounts to [2\%,12.4\%], with all values being higher than zero. Moreover, this estimate relies mainly on control units with an upwards demand trend. Therefore, we conclude that this estimate serves as a lower bound of the effect. Summing up, our paper provides the first empirical evidence, at least for Geneva, that a fare-reduction policy can help increase passenger demand. Finally, note that such quasi-experimental evidence is crucial, as price elasticities are often based on Stated Preference or experimental surveys \citep[in Switzerland, see, e.g.,][]{Weis2016SpBefragung,axhausen2021empirical}. 

The rest of this paper proceeds as follows. Section \ref{Litrev} discusses the existing literature on pricing policies in public transportation. Section \ref{Background} describes the institutional background of the Geneva case study. In Section \ref{SCM}, we discuss the methodology and our strategy to identify the estimate of interest. Moreover, we present the underlying assumptions of our so-called natural experiment. Section \ref{data} describes our unique data set, derived from the annual reports of Swiss transport companies and discusses our proposed aggregate metric. Section \ref{results} applies the synthetic control method to our case and discusses the robustness of our results. In Section \ref{Discussion}, we discuss our estimates by arguing how to achieve at a lower bound. Moreover, we debate about the so-called external validity. That is the ability of our study to produce an effect of the price reduction on demand, our theoretical mechanism of interest, to work in public transport settings. Section \ref{Conclusion} concludes.
	
\section{Literature review}\label{Litrev}

Our study fits into the literature on fare-policy interventions in urban public transport systems. \cite{bresson2003main} suggest that demand is less sensitive to fare changes in France's urban areas than in non-urban areas of England. Moreover, \cite{bresson2004economic} analyse French urban areas in greater depth and show that the effects of changing fares vary across areas. This heterogeneity is mainly explained by car ownership, urban sprawl, and the aging of the population. Recently, \cite{kholodov2021public} estimate the effect of a new fare policy in Stockholm and find varying effects across socioeconomic groups and different modes of public transport. Many other studies have examined fare policies in European cities by simulating fare changes (e.g., \cite{parry2009should} for London or \cite{matas2020economic} for Barcelona) or by analysing transport policy bundles \citep[e.g.,][for Vienna]{buehler2017vienna}. We add to such studies by calculating the causal effect of fare-policy intervention in the interesting case in Geneva.

In the literature, causal analysis has mainly been conducted on fare-free policies rather than fare reductions, as in our case. In Europe, \cite{cats2017prospects} suggest that free fares increased public transport use by 14\%. In addition, \cite{de2006impact} and \cite{rotaris2014impact} investigate free-fare policies in Brussel and Trieste. The settings of \cite{lee2019causal} in Taichung (Taiwan) and \cite{shin2021exploring} in Seoul (South Korea) are the closest to ours. In Taichung, bus network and schedule improvements gradually increased bus use, which then grew further due to free-fare policies, leading to further adjustments on the supply side. \cite{shin2021exploring} estimates there was a 16\% increase in subway use by older adults after a fare-free policy was introduced for this age group in Seoul. 

Our study is also related to the rich literature on price elasticities in public transport. Price elasticities show the percentage change in demand due to a one percentage price change. For example, \cite{holmgren2007meta} exposes a short-run price elasticity of -0.75 and a long-run price elasticity of -0.91 in Europe. In line with \cite{holmgren2007meta}, \cite{brechan2017effect} finds that increasing frequencies has a higher elasticity than reducing fares for public transport. \cite{wardman2018review} show that the effects of price changes in public transport on car demand – the so-called cross-elasticities – are relatively low. \cite{liu2019evaluating} add that changes to fare policy in Australia mostly increased the number of trips of existing users rather than attracted new users. That is why \cite{litman2004transit} suggests a relatively large fare reduction is crucial for car-users to switch to public transport. \cite{redman2013quality} show that price can encourage car-users to use public transport. However, the reliability, frequency, and speed of public transport will determine whether their intentions are implemented and maintained. In Switzerland, where our case study of Geneva is located, price elasticities regarding the demand for public transport are typically low according to \cite{CitecIngenieurs2021}. In a recent experimental study, \cite{axhausen2021empirical} estimate a price elasticity of -0.31 in Switzerland.

More broadly, our study adds to the literature on price policies, inter alia with the goal of making mobility more sustainable. For instance, \cite{kilani2014road} show that road-pricing combined with higher public transport fares in peak periods or discounts on off-peak tickets work in complementary fashion in Paris. Moreover, the effect of road-pricing \citep[e.g.,][for Milan]{percoco2015heterogeneity} or peak-pricing (off-peak discounts) in public transportation alone is also analysed in recent literature (see, e.g., \cite{rantzien2014peak} for Stockholm and \cite{huber2021business} for Switzerland). \cite{gkritza2011estimating} assess the multimodal context of the urban public transport system with varying fare structures in Athens. For a review of public transport policies, see also \cite{horcher2021review}.

Finally, we add to transportation studies applying the synthetic control method, according to \cite{athey2017state}, \textit{"the most important innovation in the policy evaluation literature in the last 15 years"} (p. 9). For instance, \cite{percoco2015heterogeneity}, also previously mentioned, investigates the effect of road-pricing on traffic flows. Another example is \cite{tveter2017fixed}, who evaluate which transportation projects affect settlement patterns. \cite{doerr2020new} estimate the extent to which new airport infrastructure promotes tourism. Studying ski-lift companies, \cite{wallimann2020complementary} discusses the effect of radically discounting prices, while \cite{xin2021impacts} investigate the impact of COVID-19 on urban rail-transit ridership.
	
\section{Background}\label{Background}

Switzerland is densely populated and has one of the highest GDP per capita in the world.\footnote{See \url{https://data.worldbank.org/indicator/NY.GDP.PCAP.CD} (accessed on November 9, 2021) } The road and rail infrastructures are modern and well maintained. Public transport is reliable and frequent, and the tariff system is widely integrated. The mixture of short distances, high incomes and good quality drives the demand for mobility in Switzerland. For these reasons, the countries' residents are highly mobile. On the one hand, 1,000 residents own, on average, about 500 individual motorized vehicles.\footnote{See \url{https://www.bfs.admin.ch/bfs/en/home/statistics/mobility-transport.html} (accessed on November 9, 2021)} Apart from a yearly fee of 40 Swiss francs to use the highways, roads are free of charge. On the other hand, every second resident owns a public transport pass.\footnote{See \url{https://www.bfs.admin.ch/bfs/en/home/statistics/mobility-transport.html} (accessed on November 9, 2021).} For example, about 2.7 million individuals (roughly 32\% of the population) held a half-fare travel ticket in 2019.\footnote{See \url{https://reporting.sbb.ch/verkehr} (accessed on November 9, 2021). Moreover, all under 16 years old (roughly 16\% of the population) also travel with a price reduction of 50\% and therefore do not need half-fare travel tickets. } With such a half-fare travel ticket, a person can buy Swiss-wide public transport tickets on a reduced tariff of 50\%. Furthermore, Swiss residents bought more than one million subscriptions to regional tariff associations in 2019 \citep{VoeV2020}. 

Switzerland is organized into 26 federal states, the so-called cantons. In Geneva, our canton of interest, 27.6\% of the residents own a Swiss-wide public transport subscription. This is relatively low compared to other Swiss agglomerations. On the other hand, the proportion with a subscription from the regional tariff association is in Geneva rather large with 25.4\% compared to other Swiss agglomerations \citep{FederalStatsOffice2010}. That is probably because of the urbanity and the small size of the canton of Geneva. For example, most journeys related to work are made within the canton \citep{FederalStatsOffice2010}.

Besides the federal system, the Swiss political system is a direct democracy. Therefore, electorates can decide on political issues at the communal, cantonal and federal state levels. In this political framework, the electorates of the canton of Geneva chose to reduce the prices of state public transport in 2013. This initiative originated from a senior citizens' association. At the request of Geneva’s population, the tariff association in Geneva implemented a sharp price reduction in December 2014. First, the full-fare hourly tickets were reduced by 14.3\% and the corresponding half-fare tickets by 20\%. Second, the full-fare daily tickets were discounted by 5.7\% and the corresponding half-fare tickets by 3.9\%. Third, adults benefited from a price reduction of 28.6\% on annual season tickets and seniors and juniors (people between 6 and 24 years) from a price reduction of 20\% and 11\%, respectively. Fourth, seniors additionally received a 10\% discount on monthly season tickets, whereas adults and juniors received no discounts on monthly season tickets \citep{UniresoGeschaeftsbericht15}. In 2014, the ticket categories who received a discount made up 65\% of the total traffic revenue for 2nd class tickets.\footnote{2nd class tickets account for almost the entire revenue.} Considering the revenue shares per ticket category of 2014, we assess an overall price discount of 12.6\%. In Appendix \ref{Appendix_A} we describe how we calculate this price change. Note that we do not account for substitution between ticket categories, which might lead us to an underestimation of the reduction. 

In summary, the policy intervention in December 2014 was the largest price reduction in a long time. The annual season ticket in Geneva now costs 500 Swiss francs for adults (previously 700 Swiss francs) and 400 Swiss francs for seniors and juniors (previously 500 and 450 Swiss francs respectively). These prices are more than 200 Swiss francs less than those charged by other Swiss cities. For instance, annual season tickets in Lausanne, Berne, Basel, and Zurich cost 740, 790, 800 and 782 Swiss francs respectively. The same is the case for single fare tickets amounting to 3 Swiss francs in Geneva. This shift away from the typical price level in Switzerland is the point of departure for our analysis. Using real-world data, we measure the effect of the price reduction on demand for public transport by comparing Geneva, where the political intervention occurred, with other regions of Switzerland.

In 2014, most of Geneva's tariff associations' revenue stemmed from TPG, a transportation company that transported about 197.1 million passengers that year.\footnote{TPG also operates to a small extent outside the regional tariff association of Geneva (also outside Switzerland).} The demand increased to 200.3 million passengers in 2015, which is an increase of 1.5\% compared to 2014. From 2014 to 2015, TPG’s traffic revenue fell from 153.7 million to 142.6 million Swiss francs \citep{TPGGeschaeftsbericht15}. We present the annual traffic revenue of TPG in Table \ref{trafficrevenue} in the Appendix \ref{Appendix_B}. The TPG transport system depended on buses in the 20th century \citep{fitzroy1999season}. However, at the beginning of the 21st century, TPG started to expand its tram network, which grew from 14.5 kilometres in 2005 to 33 kilometres in 2012 \citep{TPGGeschaeftsbericht12}. Overall, the number of vehicle kilometres increased from about 20 million in 2005 to about 29 million in 2013. Besides TPG, Geneva's tariff association consists of the Swiss Federal Railways, operating on the regional railway network, and Mouettes, which runs ferries. 

\section{Synthetic control method}\label{SCM}
In this section, we outline the synthetic control method used in our empirical analysis. Second, we present the assumptions underlying our analysis. 

\subsection{Methodology and implementation}\label{MethAndImpl}
Let $D$ denote the binary treatment 'price reduction' and $Y$ the outcome 'public transport demand'. The treatment $D$, the result of the initiative in Geneva, affects one unit (TPG). All the other units (transport companies) in our data are not exposed to the price reduction and thus constitute the control group. We can define the observed outcome of TPG, our unit of interest, as
\begin{equation}
	Y_{t}=Y_{t}^N+\alpha_{t} D_{t}.
\end{equation}

$Y_{t}$ denotes the observed outcome, $Y_{t}^{N}$ the outcome without the treatment, and $\alpha_{t}$ the treatment effect at time $t$. It is important to note that the treatment $D$ takes the value 0 for all units during the period $t < T_0$, with $T_0$ indicating the introduction of the treatment. This is because also TPG was not exposed to the price reduction during the pre-treatment period. Only looking at the post-treatment period permits to define the treatment effect as
\begin{equation}
	\alpha_{t}=Y_{t}-Y_{t}^N.
\end{equation}

As we observe $Y_{t}$, we merely need to estimate $Y_{t}^N$, the public transport demand of TPG without the policy intervention. Using statistical parlance, $Y_{t}^N$ is a counterfactual. That is the outcome one would expect if the intervention had not been implemented. 

To determine $Y_{t}^N$, we use the synthetic control method of \cite{abadie2003economic} and \cite{abadie2010synthetic}. To construct the synthetic control unit ($Y_{t}^N$), the synthetic control method uses a data-driven procedure. In our study, the counterfactual $Y_{t}^N$, the synthetic TPG, is created out of already-existing companies of the control group, the so-called 'donor pool'. For this purpose, the methodology assigns a weight to each transport company in the control group. These weights are non-negative and sum up to one. On the one hand, we assign large weights to companies with a sizeable predictive power for TPG. On the other hand, transport companies in the control group with a low predictive power receive a small or a zero weight. The goal is to minimize the difference between TPG and the synthetic TPG in the period $t < T_0$, the pre-treatment period. To discuss the success of this goal, we calculate the mean squared prediction error (MSPE) of the outcome variable between TPG and the synthetic TPG. 

To evaluate the significance of the results, we run placebo studies. To this end, we apply the synthetic control method to one transport company after another in the control group, all known to be untreated, using the remaining control companies as the donor pool. More precisely, we iteratively estimate placebo estimates of each unit with no price reduction considering it to be 'pseudo-treated'. If the estimated effect for TPG is similar to the placebo estimates, our result could have happened by chance. However, suppose the placebo investigations show that the effect estimated for TPG is enormous relative to the transport companies in the control group. In that case, like \cite{abadie2010synthetic}, we interpret our analysis as providing significant estimates of the treatment effect $\alpha_{t}$. In implementing the synthetic control method, we use the \textit{synth} and \textit{SCtools} packages for the statistical software \textsf{R} by \cite{hainmueller2015synthPackage} and \cite{PackageSCtools} respectively.

Moreover, we calculate the corresponding 95\% bootstrap confidence intervals to the average treatment effect. Therefore, we randomly draw control units with replacement from our donor pool 2,000 times to arrive at these confidence intervals. In every sample, we construct a synthetic TPG and estimate the average gap between TPG and its counterfactual.  

\subsection{Assumptions}\label{Assumptions}
Identification requires statistical procedures, as explained in the previous chapter. However, on the other hand, ensuring that our calculation identifies the effect of the price reduction also relies on assumptions about how the world, here the world of public transportation, works \citep[see, e.g.,][]{huntington2021effect}. Therefore, in this section, we discuss the assumptions underlying our analysis. 
\newline
\textbf{Assumption 1 (no anticipation):} \newline
Assumption 1 is satisfied when the public transport demand in Geneva did not change due to forward-looking customers reacting in advance to the policy intervention. To this end, the price reduction effect would be biased if TPG's travelers already use public transport before the intervention because they know that prices will fall later. \newline
\textbf{Assumption 2 (availability of a comparison group):} \newline
By Assumption 2, there exists a donor pool. The assumption is satisfied when we have a control group with characteristics that are, by assumption, comparable to the treated unit. That implies that other public transport companies do not sharply lower fares in our natural experiment. \newline
\textbf{Assumption 3 (convex hull condition):} \newline
Assumption 3 is satisfied when pre-treatment outcomes of the synthetic counterfactual can approximate the outcomes of the treated unit. Using statistical parlance, the pre-treatment outcomes of the treated unit are not ’too extreme’ (too high or low) compared to the outcomes of the donor pool.
\newline
\textbf{Assumption 4 (no spillover effects):} \newline
Assumption 4 is fulfilled when the price reduction has no spillover effects, eighter positive or negative, on other transport companies in the donor pool. An obvious failure of this assumption would be a decrease in public transport demand of other Swiss cities because their residents perceive the ticket costs as too high after the price reduction in Geneva.  
\newline
\textbf{Assumption 5 (no external shocks):} \newline
Applying the synthetic control method, we assume that no shocks occur to the outcome of interest during the study period \citep[see, e.g.,][]{abadie2021using}.  In our case, this condition is challenging, since public transport companies expand the network from time to time, which typically affects the demand for public transport \citep[see, e.g.,][]{brechan2017effect,holmgren2007meta}. To account for such changes in supply, we propose an aggregate metric that breaks down the demand for public transport per company's supply, which we use as our outcome variable. More precisely, we calculate the ratio of passenger trips per vehicle kilometre, being robust against changes on the supply side.

\section{Data}\label{data}

To investigate the effect of the policy intervention in Geneva, we use the annual reports of Swiss transport companies, which the Swiss National Library systematically archives.\footnote{See \url{https://www.nb.admin.ch/snl/de/home.html} (accessed on November 9, 2021). In this study, we focus on transport companies, as annual reports are not publicly available for tariff associations in the period of interest. } In these annual reports, the companies publish financial and non-financial performance indicators. We systematically gathered the most relevant performance indicators from public transport companies for our dataset. TPG operates mainly in the city of Geneva, the densest and second largest city in Switzerland, and its agglomeration belt. Using the synthetic control method, we have to choose each unit in the donor pool judiciously to provide a reasonable control for TPG, the treated unit (see Assumption 2 in Section \ref{Assumptions}). Therefore, we only consider transport companies that operate trams and buses primarily in cities with more than 50,000 inhabitants. These are Bernmobil (Berne), BVB (Basel), SBW (Winterthur), TL (Lausanne), TPL (Lugano), VB (Biel), VBL (Lucerne), VBSG (St Gallen) and VBZ (Zurich).\footnote{The VBL (Lucerne) provided us with the VBL data, as it was not publicly available. }  

First, we collected the number of passenger trips, which are standardized in Switzerland. The number of passenger trips counts how many passengers enter a company’s vehicle per year. Passenger trips are essential, as we want to measure the increase in public transportation use, which, e.g., could be due to a mode shift from car use to mass transportation. Today, companies mainly count passengers automatically, but this was often done by hand in the past. This change of the counting system happened in Geneva from the years 2015 to 2016. Therefore, we adjust our TPG data from 2016 to 2019 based on the observed growth rate of the passenger trips to have a uniform panel dataset.\footnote{We have verified our adjustments with the transport company TPG.} Since 2005, TPG has experienced the highest increase in passenger trips (compared to Swiss transport companies in the donor pool), followed by TL operating in Lausanne, another city in the French-speaking part of Switzerland. However, since 2005, TPG, together with VBSG (St Gallen), has also experienced the highest increase in vehicle kilometres. The increase results from the extension of tram routes. Therefore, to mitigate changes in supply, i.e., external shocks increasing company's networks (see Assumption 5 in Section \ref{Assumptions}), we use the previously discussed aggregate metric of passenger trips per vehicle kilometre as the outcome variable. 

\begin{figure}[H] 
		\centering \caption{Ratio \label{Ratio}}
		\includegraphics[scale=.6]{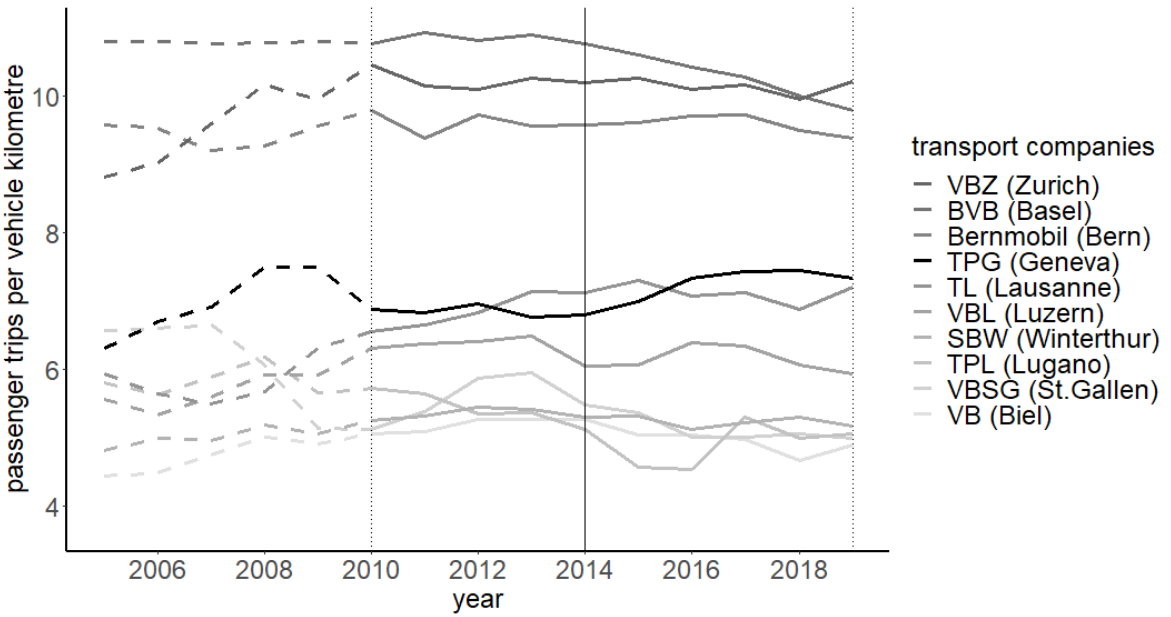}
	\end{figure}

Consequently, companies with high-capacity utilization would have a high value in our dependent variable. Figure \ref{Ratio} shows the development of passengers per vehicle kilometre for companies operating in a city with more than 50,000 inhabitants over time. As expected, our metric is mainly robust against changes on the supply side. Moreover, this variable also serves as a proxy for an average passenger load rate. Considering CO\textsubscript{2} emissions, this is also important, as the average emissions of each individual passenger decrease when the metric increases. Finally, we also observe that TPG is not extreme in the values of the outcome variable before the intervention. This is important to define a weighted subset of control companies that is comparable to TPG (see Assumption 3 in Section \ref{Assumptions}). 
	
Controlling for supply changes also makes sense, as several studies show considerable effects of vehicle kilometers on demand \citep[see, for instance,][]{holmgren2007meta}. In other words, and as a word of caution, we assume a considerable supply elasticity when applying the ratio. However, and also a thing to notice in Figure \ref{Ratio} by looking at the period with the dotted lines, due to a substantial increase in vehicle kilometers plied by bus lines in Geneva’s agglomeration belt from 2008 to 2010, the ratio in Geneva declined. This is because the aggregate change in TPG's supply occurred in the subarea where public transport is relatively poorly utilized. Therefore, we restrict our pre-treatment period to the years 2010 to 2014. However, collecting several observations on the unit of interest (TPG) and the donor pool is crucial before the price reduction \citep{abadie2021using}. Therefore, we also perform a robustness check with a more extended pre-treatment period. Moreover, we also oppose our results to estimations without the metric and thus use only passenger trips as the outcome variable. This robustness check is crucial, as unexpected low (or high) supply elasticities could be an alternate explanation of the treatment effect.

We match our outcome variable with predictors, forces working in a public transportation setting, to predict our outcome variable and build a valid synthetic TPG. We, therefore, gathered aggregate data about the share of public transport and individual motorized vehicles in Swiss urban areas from the Swiss Mobility and Transport Microcensus for 2010\footnote{See \url{https://www.bfs.admin.ch/asset/de/su-d-11.04.03-MZ-2010-G07.3.1.1} (accessed November 9, 2021).} and 2015\footnote{See \url{https://www.bfs.admin.ch/bfs/en/home/statistics/mobility-transport.html} (accessed November 9, 2021).}.  We use these modal-split predictors to map the choice of transport modes in each urban area. In addition, we include variables for population growth and population density yearly provided by the association of cities.\footnote{See \url{https://staedteverband.ch/de/Info/publikationen/statistik-der-schweizer-stadte} (accessed October 21, 2021).} These variables account for the potential demand for public transport in a given region. Finally, we use the average of pre-treatment outcomes for 2012 to 2014 (after 2011, the tram network of Geneva did not expand any further) for treated and control units as predictors.

\section{Results}\label{results}

We subsequently present the results of applying the synthetic control method, evaluate their significance and investigate their robustness.

\subsection{The effect of the price reduction}\label{effect}

\begin{figure}[H] 
		\centering \caption{Demand development of TPG and the synthetic TPG \label{Demand_Dev} }
		\includegraphics[scale=.6]{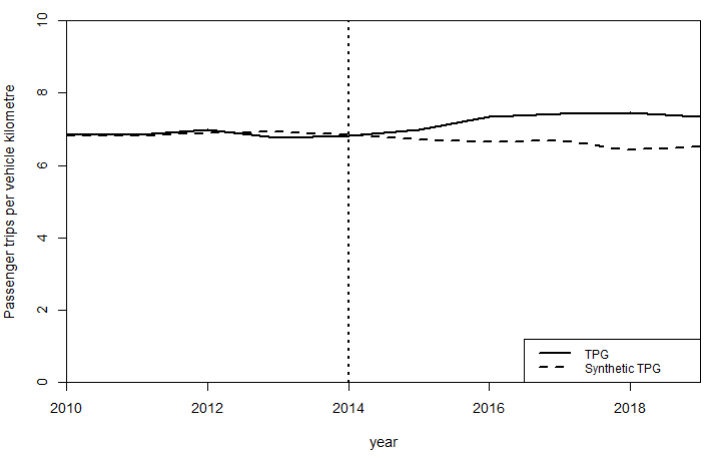}
	\end{figure}

To construct the synthetic TPG, the synthetic control method assigns weights among the control group companies. VB (Biel) receives the highest weight with 0.400, while BVB (Basel) has the second-highest weight with 0.162, and the VBSG (St Gallen) has a zero weight. Table \ref{Weights} in Appendix \ref{Appendix_C} shows the weights for each company in the donor pool. Figure \ref{Demand_Dev} plots the outcome variable, equal to passenger trips per vehicle kilometre, of TPG and the synthetic TPG from 2010 to 2019. We can easily observe that the two trajectories track each other close in the pre-treatment period, i.e., the pre-price-reduction period. Thus, the mean squared prediction error (MSPE) of the outcome variable between TPG and the synthetic TPG amounts to a small figure of 0.009. Therefore, our synthetic TPG is a sensible counterfactual of the outcome we would expect if the intervention had not been implemented. While demand from customers of the synthetic TPG continued its slightly downward trend, the demand for TPG increased. This difference is relatively constant over four years, from 2016 to 2019. 
	
The estimate of our analysis indicates the effect of the policy intervention on demand in passenger trips per vehicle kilometre. More precisely, after the price reduction, this effect represents the yearly differences (gaps) between TPG and its synthetic counterfactual. On average, the demand (our ratio) increased by about 0.72 from 2015 to 2019. In other words, almost one additional passenger per vehicle kilometre boarded TPG’s buses and trams due to the price reduction, an increase of about 10.6\% compared to 2014. Thus, we conclude that we can infer a positive effect on demand in Geneva due to the price reductions. Randomly drawing nine control units with a replacement from our donor pool leads us to bootstrap confidence intervals. The corresponding 95\% bootstrap confidence interval of the average estimated effect is [0.423; 0.870]. We present the distribution of the means of 2,000 samples in Figure \ref{Bootstrap} in Appendix \ref{Appendix_C}. 

The black line in Figure \ref{Placebos} illustrates the gap between the trajectories of TPG and the synthetic TPG. As we know from the results above, the MSPE of the outcome variable between TPG and the synthetic TPG is small. Hence the trajectories track each other closely in the pre-treatment period. However, they separate in the post-treatment period, and therefore we observe a causal effect of the treatment (price reduction) on demand (aggregate metric). We can now construct a synthetic counterfactual for all companies in our control group and compare these trajectories to the actual company's development. Suppose the trajectories from the 'pseudo-treated' companies and their synthetic counterpart fit well in the pre-treatment period and separate in the post-treatment period (even though they have not introduced a price reduction). Then, our effects calculated for TPG may be caused by chance rather than by the treatment (the price reduction). 

	\begin{figure}[H] 
		\centering \caption{Demand gaps of TPG and control companies \label{Placebos}}
		\includegraphics[scale=.4]{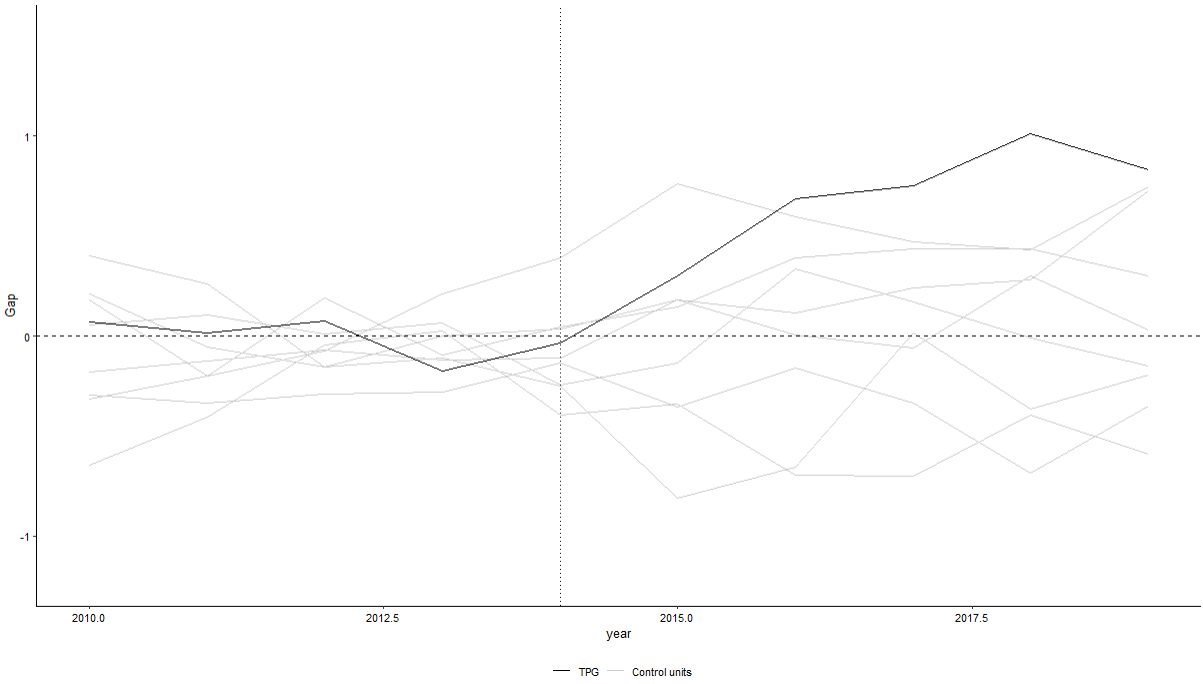}
	\end{figure}

The grey lines in Figure \ref{Placebos} summarize the results of iteratively applying our method to one transport company after the other by illustrating the gaps between the actual and the synthetic trajectories. The average MSPE among the companies amounts to 0.09, and the median amounts to 0.07, figures which are relatively small. Hence the trajectories track each other closely in the pre-treatment period. In other words, the methodology also provides mainly suitable counterfactuals for most companies in the control group. However, there remain a few lines that still deviate substantially from a zero-gap. From 2016 to 2019, the black line, the gap between TPG and its synthetic counterfactual, is further apart than all the grey lines. Hence, the difference between the post-treatment MSPE and the pre-treatment period is the greatest among the companies. The ratio for TPG amounts to 66.0, while the companies with the second and third highest ratios are VBZ (Zurich) and TL (Lausanne), with 9.7 and 5.5 respectively. On average, the post-treatment MSPE divided by the pre-treatment MSPE amounts to 3.7 in the donor pool. Therefore, we conclude that the increase in demand for TPG due to the price reduction is not driven by chance. 

\subsection{Robustness analysis}\label{robust}

In this section, we challenge our assumptions and our study design by performing robustness investigations. First, as a methodological robustness check, we apply a recent development of the synthetic control method, the synthetic difference in differences approach of  \cite{arkhangelsky2019synthetic}, to demonstrate the goodness of our results. Second, we expand our pre-treatment period. Third, we expand our donor pool with companies operating in cities with fewer than 50,000 inhabitants. Fourth, we estimate the effect of the lower fares on the number of passengers (and not the number of passengers per vehicle kilometre). In the fourth robustness check, the synthetic TPG does not mimic TPG in the pre-treatment period appropriately. Therefore, in a final robustness investigation, we re-estimate the effect on the number of passengers using the synthetic difference in differences approach.

As a first robustness check, we use the synthetic difference in differences approach proposed by \cite{arkhangelsky2019synthetic}. In a nutshell, this methodology decides in a data-driven way (through minimization of the MSPE in pre-treatment periods) whether the synthetic control methodology or the conventional difference in difference model \citep[see, e.g.,][]{card2000minimum} is more appropriate for a case. However, this is a simplification. More precisely, it is an extension of the synthetic control method because it includes unit-specific fixed effects control for constant differences in the demand level among the affected unit and the synthetic counterfactual. Hence, the demand level can vary by a constant. Therefore, it might be sufficient that the affected unit and the synthetic counterpart match each other in terms of changes or trends rather than levels (as the difference in difference model might be more appropriate). Moreover, the synthetic differences-in-differences method includes time weights. By weighting each period before the intervention, the synthetic difference-in-difference method is more sensitive to what is happening in the periods just before the price intervention. We use the \textit{synthdid} package by \cite{arkhangelsky2019synthetic} to implement the synthetic difference in differences method. Almost identical to our original result, the demand increases by 0.68, or 10.0\%. Therefore, we can draw the conclusion that the original synthetic control method is appropriate for our case.

Due to a significant increase in the vehicle kilometres of bus lines in Geneva's agglomeration belt from 2008 to 2010, we restricted our pre-treatment period to 2010 to 2014. However, it is crucial when applying the synthetic control method not to have a pre-intervention window that is too small. Therefore, in a second robustness check, we substantially expand our pre-treatment period to 2005 to 2014. In this analysis, the MSPE amounts to 0.1, which is slightly worse (but still decent) compared to our original estimation. The estimate amounts to 0.61, indicating an increase in demand of about 9\% for TPG (compared to 2014). That is about 1.5 percentage points lower than the main result and is, therefore, almost identical to our original result.

Due to the risk of over-fitting, we only include transport companies operating in cities with more than 50,000 inhabitants in the control group. However, as the design of our donor pool might influence our results, we expand the donor pool in a third robustness check with transport companies from smaller cities that also primarily operate trams and buses and for which the necessary data are available. These are BBA (Aarau), BSU (Solothurn), MBC (Morges), STI (Thun), TPN (Nyon), Travys (Yverdon), VBG (Zurich agglomeration), VZO (Zurich agglomeration) and ZVB (Zug). The estimate amounts to 0.71, an increase in demand of 10.5\% from 2015 to 2019 (compared to 2014). The pre-treatment mean squared prediction error (MSPE) of the outcome variable between TPG and the synthetic TPG is impressively low, amounting to 0.09. Therefore, we conclude that we construct a decent counterfactual and the estimate of our original study design is robust. 

In a fourth robustness investigation, we replace our metric with the original number of passenger trips. Remember, due to variation in vehicle kilometres of trams in Geneva in the study period and their effect on passenger trips, we define the ratio of passenger trips per vehicle kilometre as our outcome variable. When we analyse the impact on passenger trips alone, we arrive at similar patterns but a higher MSPE. Therefore, the trajectories of TPG and the synthetic TPG do not track each other as closely in the pre-intervention period as in our main study design (see also Figure \ref{Demand_Dev_Einsteiger} in Appendix \ref{Appendix_C}). The reason for this is the positive demand shock in 2012 due to the finalization of the tram network extension. The estimate, however, is comparable amounting, to 18.0 million additional passenger trips. That points to a demand increase of about 9.1\% for TPG compared to 2014, which is 1.4 percentage points lower than the main result. Note that these figures should be interpreted with much caution, as the pre-treatment fit is not decent, and TPG already starts (at 2014) at a higher value than the synthetic TPG. Moreover, an important thing to notice is that the outcome of TPG (passenger trips) is already higher than the synthetic counterpart (see Figure \ref{Demand_Dev_Einsteiger}). Therefore, it might be that the effect due to the price reduction is lower than reported. 

Adding to the fourth robustness check, we again estimate the price reduction effect on passenger trips. However, different from the previous investigation, we use the synthetic difference in differences approach instead of the synthetic control method. The recent methodology might improve as it is more sensitive to what is happening in the periods just before the intervention. Moreover, it permits the outcomes of TPG and the synthetic TPG to differ as it includes unit fixed effects. We find that the effect amounts to, on average, 7.3 million additional passenger trips. That is an increase of about 3.7\% for TPG compared to 2014, which is lower than our original result. Figure \ref{SDID_Passengers} in Appendix \ref{Appendix_C} plots the effect estimated with the synthetic difference in differences approach. In our case, the methodology focuses on the parallel trend between TPG and TL (Lausanne), the company with the most favorable demand development among the unaffected transport companies. However, note that the corresponding 95\% bootstrap confidence interval of the average estimated effect of the fifth robustness check is [3.9 million; 24.5 million]. That is an increase between 2.0\% and 12.4\%, with all values being higher than zero. 

\section{Discussion}\label{Discussion}
We assess a demand the effects of lower urban public transport fares and find that the price reduction in Geneva leads to a demand increase of about 10.6\%. To isolate the effect of our mechanism of interest, the price reduction, we propose an aggregate metric inheriting supply changes of public transport networks. This makes sense as we are able to block off the effect of increasing and decreasing frequencies as an alternate explanation of demand-effects, being in the context of public transport of crucial importance. Moreover, robustness investigations show that the estimate is robust when we modify the study design, i.e., longer pre-treatment period or more companies in the donor pool, or applying the synthetic difference in differences approach. 

However, on the other hand, the estimate is significantly lower when we consider the outcome variable passenger trips and do not isolate the price reduction effect. It amounts to 3.7\% when we apply the synthetic difference in differences methodology. This demand increase is even lower than naively comparing the passenger trips of TPG after and prior to the price discount, amounting to 5.7\% additional trips.\footnote{This increase is also due to an additional railway cross-country train line between Geneva and France, coming alongside an increase of vehicle kilometers of TPG. The latter also increases demand. However, note that we can control for this effect using our proposed aggregate metric.} This is because the estimate of the robustness check 5 mainly relies on control units with an upwards trend. Moreover, when calculating bootstrap estimates of the effect, we do not get any negative values and the 95\% bootstrap confidence interval of the average estimated effect points to an increase of between 2.0\% and 12.4\%, inclusive. Therefore, we conclude that the effect of 3.7\% additional demand is a potential lower bound of the effect. 

As the price change from 2014 to 2015 amounts to 12.6\%, we can calculate corresponding point elasticities of demand. In Appendix \ref{Appendix_A}, we describe how we assess the price elasticities. We get average elasticities of -0.84 and -0.29 of our main result and the lower bound, respectively. These estimates are in line with the literature. In particular, \cite{holmgren2007meta} proposes that the often stated rule of thumb of a price elasticity amounting to –0.3 only holds when vehicle kilometres are treated exogenous, but not when vehicle kilomteres are treated endogenously. In the latter case, \cite{holmgren2007meta} suggests a short-run price elasticity of -0.75 and a long-run price elasticity of -0.91 in Europe. 

One limitation of the study is that we did not analyze the influence of the COVID-19 outbreak. Future studies should investigate a more extended period and also take into account the impact of the pandemic. A thing to notice is that TPG is a company that operates on a cross-border territory. In Switzerland, and thus in the donor pool, we only have BVB (Basel) and TPL (Lugano) with a comparable situation. Therefore, we can not completely exclude that the price reduction has a different effect on TPG's measures than on companies in the donor pool, which might lower the external validity of our result. Moreover, it is again essential to mention the extension of the tram network in Geneva, which, as a quality improvement, could still have had after-effects on demand. Thus, using statistical jargon, we do not know whether we completely isolated the effect of the supply increase, even when applying our metric. Furthermore, future studies should aim at understanding whether a price reduction is a driver for a modal substitution, e.g., passengers switching from cars to public transport. 

Finally, note that we only present a point estimate of demand changes. Therefore, any generalizations from our findings should consider this factor. Moreover, we have considered the price reduction effect as a policy intervention and not the impact of the size of the discount. E.g., \cite{brechan2017effect} finds no significant relationship between the size of the price reduction and the demand reaction. Therefore, e.g., using similar study designs to discuss different effect sizes (if present) is on the agenda for future research. In addition, the price reduction was not the same for all age groups and ticket sentiments. Therefore, future studies could also investigate demand effects for specific client groups, e.g., seniors.

\section{Conclusion}\label{Conclusion}

In this study, we answered the question of whether a public transport price discount leads to increasing demand. Therefore, we have applied the synthetic control method to assess the demand effects of lower fares in Geneva, a Swiss urban area. The methodology is ideal for such quasi-experimental settings of price reductions (in urban areas). It constructs a counterfactual that mimics the demand a treated unit would have experienced without the price reduction in a data-driven way. Following a democratic vote, the regional tariff association in Geneva introduced a price reduction of 28\% for annual season tickets and of 20\% for hourly tickets. To the best of our knowledge, our study is the first causal analysis of this case and of price reductions due to direct democracy in general. We created a unique data set of annual reports from Swiss transport companies to identify the increase in demand. In addition, we proposed a metric for aggregate demand to block off increasing networks as an alternate explanation of demand-effects, being in the context of public transport of crucial importance \citep{brechan2017effect,holmgren2007meta}. This metric breaks down the demand for public transport per company's supply. We found that the lower fares caused an increase in demand of 10.6\% from 2015 to 2019 for TPG, by far the biggest operator in the Geneva tariff association. The result remains robust when performing several robustness checks. However, when changing study design by looking at the effect and applying the synthetic difference in differences method, we were able to provide a lower bound of the effect's estimate amounting to an increase of 3.7\% additional passenger trips.

	\newpage
	\bigskip
	
	\bibliographystyle{econometrica}
	\bibliography{Genfbib}
	
	\bigskip
	
\begin{appendix}
		
		\numberwithin{equation}{section}
		\counterwithin{figure}{section}
		\noindent \textbf{\LARGE Appendices}
	
	\section{Price elasticity}\label{Appendix_A}
	
	Taking a demand change and a price change together, we can estimate  a price elasticity of demand: 
	
	\begin{equation}
		\text{Price elasticity of demand} = \frac{\text{Demand change in \%}}{\text{Price change in \%}}
	\end{equation}
	
	To define the price change on aggregate level, we consider the revenue share of each ticket category and their price change. We calculate the price change on aggregate level, the so-called overall price change, based on the revenue share before the price intervention: 
	
	\begin{equation}
		\text{Relative price change}=\sum_{n}^{i=1}\text{revenue share}_{i,2014}*\frac{price_{i,2015}-price_{i,2014}}{price_{i,2014}}*100
	\end{equation}
	
	where $i$ denotes the ticket categories. Moreover, $1$ stays for the first an $n$ for the last category.
	
	\section{Descriptive statistics}\label{Appendix_B}
	
	\begin{table}[H]
		\begin{center}	
			\caption{Key figures of TPG and the control group (in millions)}\label{Keyfigures}
			\begin{tabular}{c|ccc|ccc}
				\hline
				\textbf{}     & \multicolumn{3}{c|}{\textbf{TPG}}                                   & \multicolumn{3}{c}{\textbf{Control group (mean)}}                   \\ \hline
				\textbf{year} & \textbf{metric} & \textbf{passengers} & \textbf{vehicle kilometers} & \textbf{metric} & \textbf{passengers} & \textbf{vehicle kilometers} \\ \hline
				2019          & 7.3             & 217.9               & 29.7                        & 7.0             & 90.4                & 10.9                        \\
				2018          & 7.4             & 210.7               & 28.3                        & 6.9             & 89.3                & 10.9                        \\
				2017          & 7.4             & 207.9               & 27.9                        & 7.1             & 89.2                & 10.6                        \\
				2016          & 7.3             & 204.5               & 27.8                        & 7.1             & 88.5                & 10.5                        \\
				2015          & 7.0             & 200.3               & 28.6                        & 7.1             & 88.5                & 10.3                        \\
				2014          & 6.8             & 197.1               & 28.9                        & 7.2             & 88.1                & 10.3                        \\
				2013          & 6.8             & 196.6               & 29.1                        & 7.4             & 88.5                & 10.2                        \\
				2012          & 7.0             & 192.3               & 27.6                        & 7.3             & 87.6                & 10.2                        \\
				2011          & 6.8             & 177.1               & 25.9                        & 7.2             & 85.3                & 10.1                        \\
				2010          & 6.9             & 172.1               & 25.0                        & 7.2             & 84.4                & 9.8                         \\ \hline
			\end{tabular}
			\par
			\textit{Note: Metric denotes our aggregate ratio being passenger trips per vehicle kilometers.}
		\end{center}
	\end{table}
	
		\begin{table}[H]
		\begin{center}	
			\caption{Traffic revenues of TPG (in millions)}\label{trafficrevenue}
			\begin{tabular}{c|c}
				\hline
				\textbf{year}     & {\textbf{TPG}}   \\ \hline
				2019&153.8\\
				2018&150.7\\
				2017&146.1\\
				2016&145.3\\
				2015&142.6\\
				2014&153.7\\
				2013&152.1\\
				2012&144.3\\
				2011&135.2\\
				2010&127.9\\  \hline

			\end{tabular}
			\par
		\end{center}
	\end{table}
	
	\section{Further tables and figures}\label{Appendix_C}
	
	\begin{table}[H]
		\begin{center}
			\caption{Company weights for the synthetic TPG}\label{Weights}
			\begin{tabular}{ll}
				\hline
				\textbf{Company} & \textbf{Weight} \\ \hline
				Bernmobil (Bern) & 0.055           \\
				BVB (Basel)      & 0.162           \\
				SBW (Winterthur) & 0.080           \\
				TL (Lausanne)    & 0.079           \\
				TPL (Lugano)     & 0.091           \\
				VB (Biel)        & 0.400           \\
				VBL (Lucerne)    & 0.083           \\
				VBSG (St Gallen) & 0.000           \\
				VBZ (Zurich)     & 0.049           \\ \hline
			\end{tabular}
		\end{center}
	\end{table}
	
	\begin{figure}[H] 
		\centering \caption{Bootstrap estimates \label{Bootstrap}}
		\includegraphics[scale=.5]{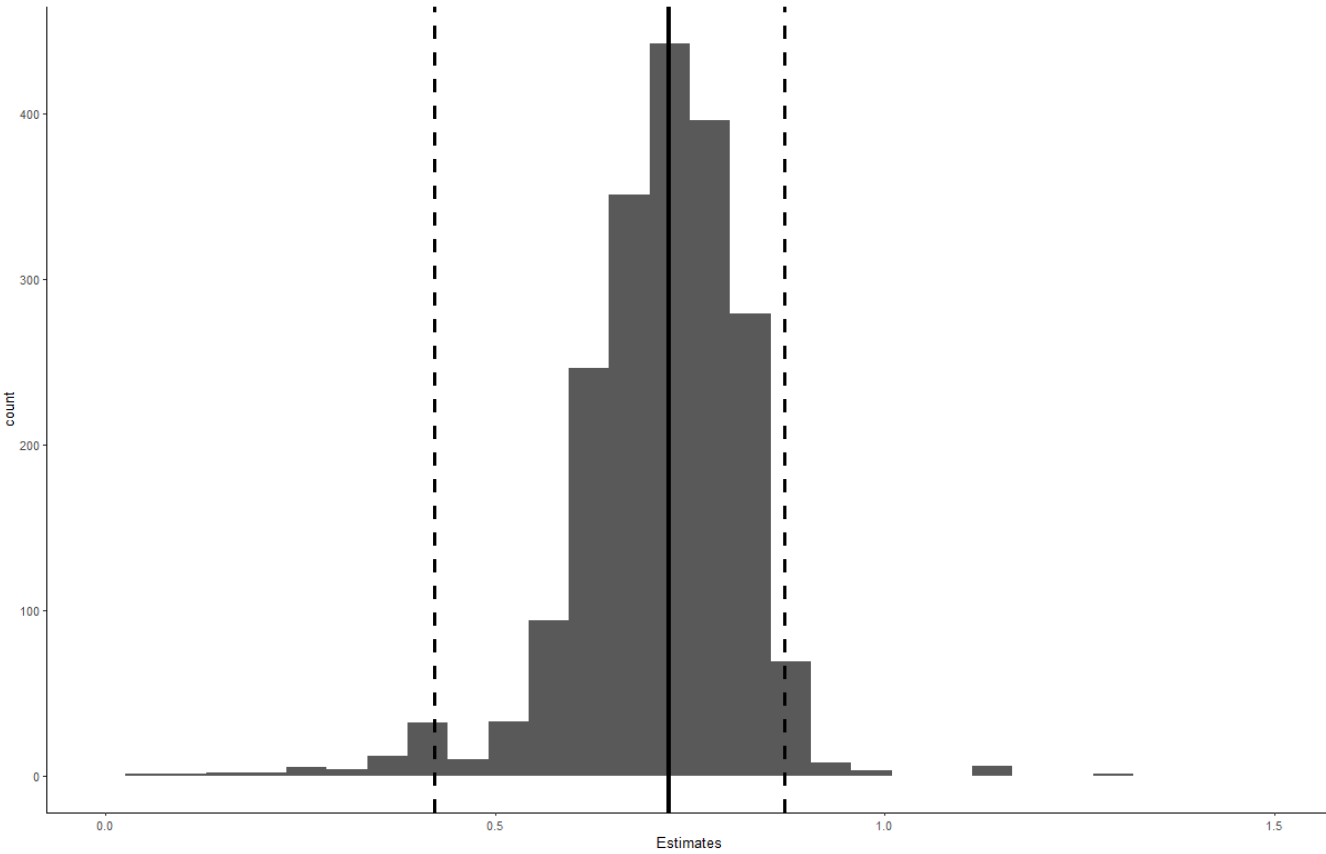}
		\par
		\textit{Note: values greater than 1.5 are not displayed in this Figure.}
	\end{figure}

	\begin{figure}[H] 
		\centering \caption{Gap between TPG and the synthetic TPG of robustness check 2 \label{Demand_Dev_Einsteiger}}
		\includegraphics[scale=.5]{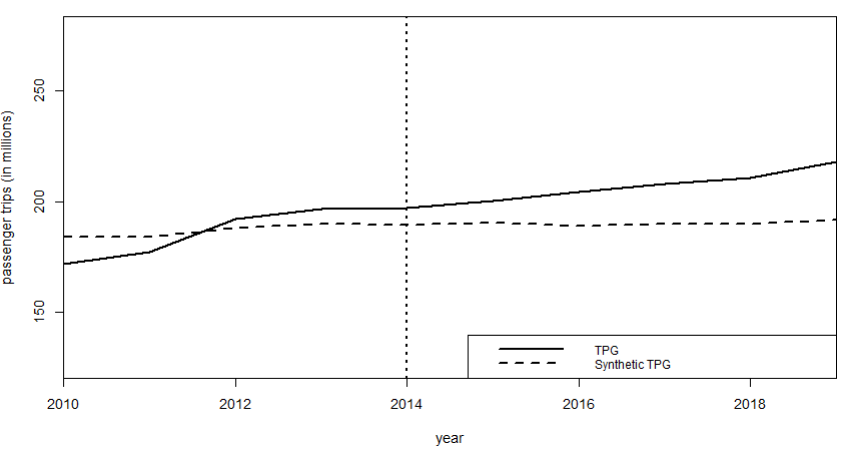}
	\end{figure}
	
		\begin{figure}[H] 
		\centering \caption{Effect on passenger trips estimated with the synthetic differenece in differences approach \label{SDID_Passengers} }
		\includegraphics[scale=.45]{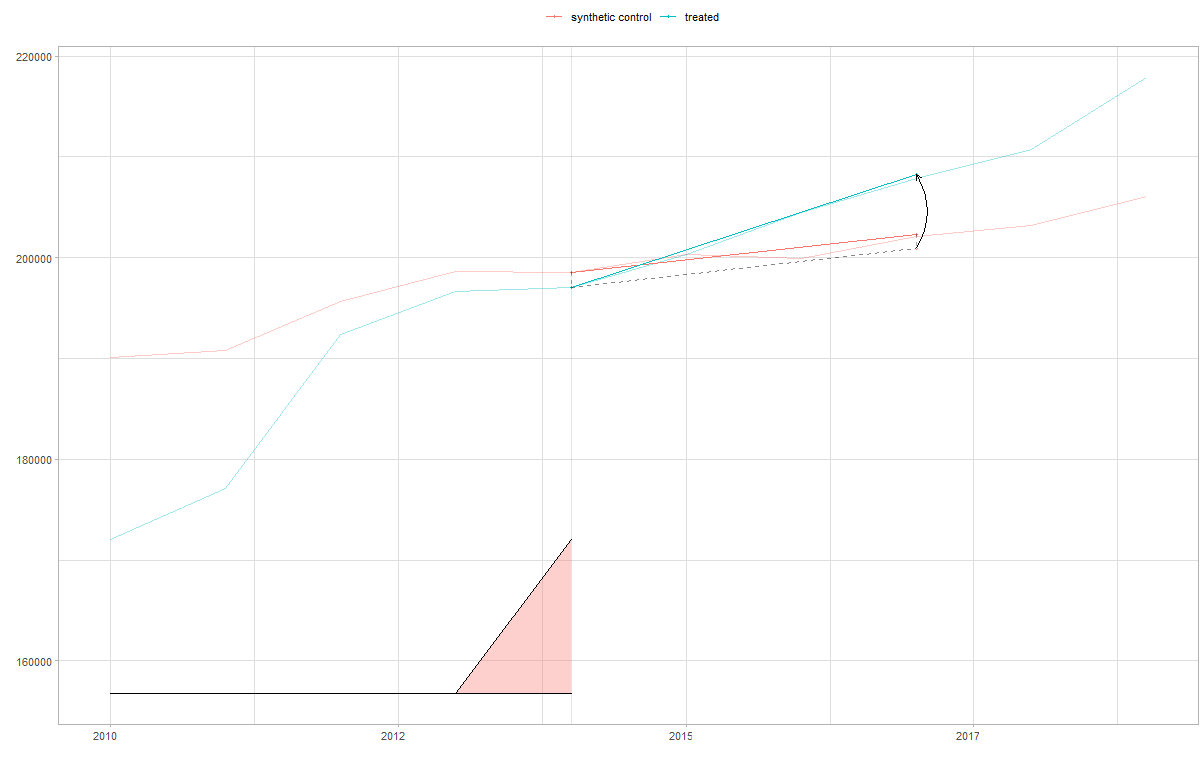}
	\end{figure}
	
		\begin{figure}[H] 
		\centering \caption{Bootstrap estimates of robustness check 5\label{Bootstrap_5}}
		\includegraphics[scale=.5]{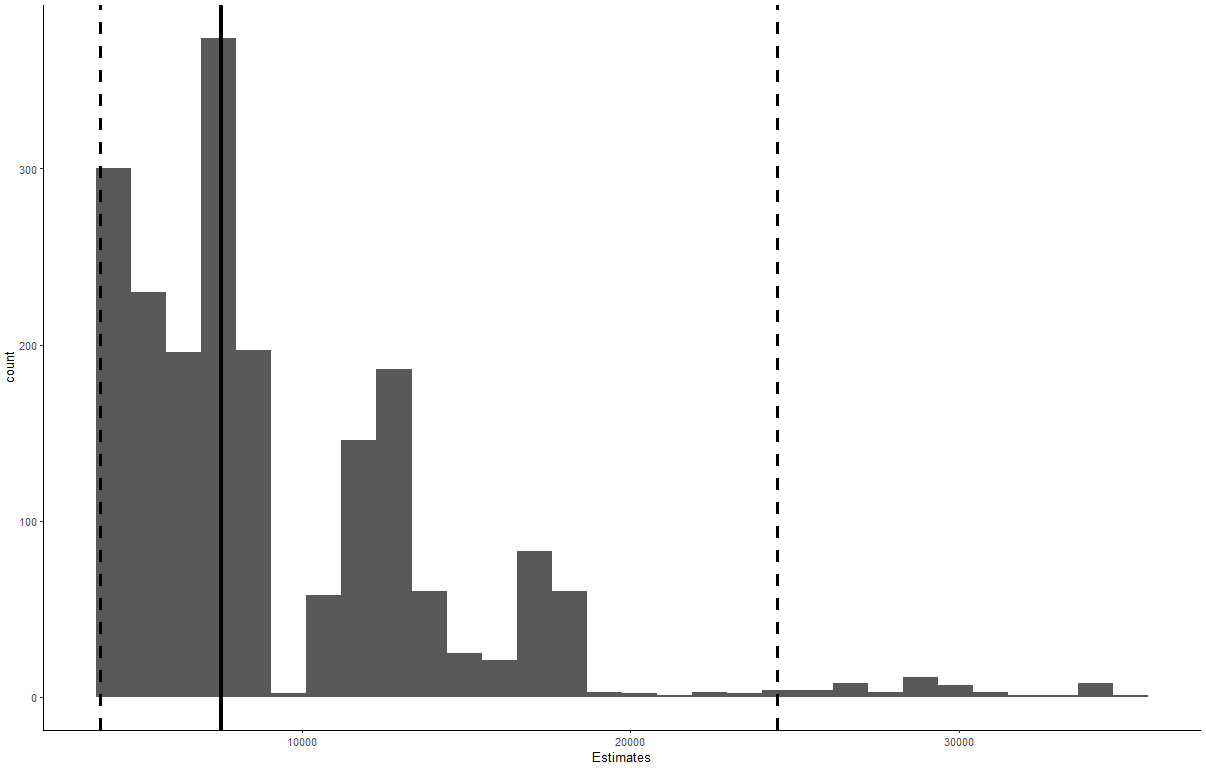}
		\par
	\end{figure}

	\end{appendix}
\end{document}